\documentclass[11pt,preprint]{aastex}
\shorttitle{Emission Corrections for Sloan Spectra}
\shortauthors{Serven et al.}
\usepackage{graphicx}
\usepackage{epstopdf}
\usepackage{float}
\begin{document}
\title{Emission Corrections for Hydrogen Features of the Graves et. al 2007 Sloan Digital Sky Survey Averages of Early Type, Non-liner Galaxies}
\author{Jedidiah Serven \& Guy Worthey}

\affil{Department of Physics and Astronomy, Washington State University,Pullman, WA 99164-2814}

\email{jdogg@wsu.edu, gworthey@wsu.edu}

\begin{abstract}
For purposes of stellar population analysis, emission corrections for Balmer series indices on the Lick index system in Sloan Digital Sky Survey (SDSS) stacked quiescent galaxy spectra are derived, along with corrections for continuum shape and gross stellar content, as a function of the Mg $b$ Lick index strength. These corrections are obtained by comparing the observed Lick index measurements of the SDSS with new observed measurements of 13 Virgo Cluster galaxies, and checked with model grids. From the H$\alpha$ Mg $b$ diagram a linear correction for the observed measurement is constructed using best fit trend lines. Corrections for H$\beta$, H$\gamma$ and H$\delta$ are constructed using stellar population models to predict continuum shape changes as a function of Mg $b$ and Balmer series emission intensities typical of H{\sc II} regions. The corrections themselves are fairly secure, but the interpretation for H$\delta$ and H$\gamma$ indices is complicated by the fact that the H$\delta$ and H$\gamma$ indices are sensitive to elemental abundances other than hydrogen.
\end{abstract}

\keywords{galaxies: abundances --- galaxies: clusters: individual: (Virgo cluster) --- methods: analytical --- surveys}

\section{Introduction}
One of the more important factors in determining the age of a stellar population such as an elliptical galaxy is being able to measure the hydrogen Balmer series \cite{wor94} and \cite{oconnell76}. The first two Balmer lines H$\alpha$ and H$\beta$ are primarily used to determine the age of a stellar population because of their relative insensitivity to changes in metallicity (\cite{serv05} and \cite{korn05}) making a measurement of these two lines a more reliable estimator of age then the higher order Balmer lines H$\gamma$ and H$\delta$, which are sensitive to changes in metallicity \cite{serv05} and \cite{korn05} making a knowledge of the relative abundances of elements in a given spectra necessary in order to get a consistent agreement of ages from all the Balmer lines.

Another important factor in determining the age of a galaxy is hydrogen emission, which can give the impression of a much older galaxy by filling in and weakening the diagnostic Balmer absorption features. Traditionally, an emission correction is used that is based on the O{\sc II} or O{\sc III} emission lines \citep{schiavon06,gonzalez93} in which the difference in the measured equivalent width(EW) of a hydrogen line such as H$\alpha$ or H$\beta$ due to emission is some fraction of the EW of O{\sc II} or O{\sc III}. However, the correlation between hydrogen and oxygen emission strengths is weak, and indeed there is no astrophysical reason why the two should be tightly correlated. On the other hand, being able to use hydrogen recombination lines, which have ratios that are nearly fixed with respect to themselves \citep{oster} should give a more reliable result.

Toward that end, we compare with 13 high-quality spectra of Virgo cluster early-type galaxies (Serven \& Worthey, in preparation) with a sample of SDSS spectra provided by G. J. Graves \citep{graves07}. 

The Virgo Cluster elliptical galaxies consist of 13 red galaxies with colors $0.75 < $B$-$V$< 0.97$. They also have a high S/N from S/N=150 to S/N=450, are well fluxed using standard star flux calibrations, and have tight control over instrument resolution and velocity dispersion. These spectra are long slit spectra obtained using the T2KB chip on the Cassegrain spectrograph on the Kitt Peak Mayall 4 meter telescope (2006 Jan 31 - Feb 5) and were chosen to cover velocity dispersions from 50 to 350 km s$^{-1}$ with a wavelength range from 3200 \AA\ to 7500 \AA\  in two wavelength swaths.

The \cite{graves07} sample was constructed by Graves from spectra of 22,501 SDSS galaxies that fall into the redshift range 0.06$< z <$0.08 and who's colors meet the criteria (g-r)$^{0.1}$ $>$ -0.25( M$_{r}$ $^{0.1}$ - 5 log h ) + 0.42, with $h = 0.70$ \cite{yan06} placing them firmly within the red sequence. These spectra are then further divided into those with H$\alpha$ and O{\sc II} emission lines and those without. Those with emission lines were said to be "LINER-like" and those without "quiescent", where LINER stands for Low Ionization Nuclear Emission-line Region.

These quiescent galaxies, which are the object of this paper, were then reduced to a sample of 2000 by constructing a Gaussian distribution centered about EW O{\sc II} = 0 \AA\  with $\sigma_{[O{\sc II}]}$ = 1.56 \AA\ to match the width of the distribution. This was then truncated at $\sigma_{[O{\sc II}]}$ to exclude outliers that may contain emission. From this distribution 2000 quiescent galaxies where randomly selected and split into 6 velocity bins such that they contain roughly equal numbers of galaxies. The bins ranges are 70$< \sigma<$120 km s$^{-1}$, 120$< \sigma < $145 km s$^{-1}$, 145$< \sigma <$165 km s$^{-1}$, 165$< \sigma < $190 km s$^{-1}$, 190$< \sigma < $220 km s$^{-1}$, and 220$< \sigma < $300 km s$^{-1}$. In order to compare the individual bins, the individual spectra within each bin were smoothed to a $\sigma$ = 300 km s$^{-1}$ and coadded to make six composite spectra. Along with these spectra the S/N at each resolution element was computed to produce an error spectrum for each of the six  composite spectra. Factors that may contribute to the error estimate are age, individual abundances, and emission signal. The error spectrum is likely to be dominated almost entirely by measurement uncertainty given the poor resolution of the individual SDSS spectra.

Measurement from these composite spectra and their errors are what is represented as the six SDSS data points in this paper's figures. For more on the binning of the SDSS spectra and more on the SDSS spectra in general see \cite{graves07}.

The central aim of this paper is to measure the strength of any residual hydrogen emission is these quiescent spectra. Noting the apparent residual H$\alpha$ emission of the SDSS spectra as seen in the first panel of Figure 1 we construct  simple formulae for the correction of the relative intensities of the rest of the Balmer series for the Sloan Digital Sky Survey (SDSS) as a function of Mg $b$ that includes continuum slope effects and the intrinsic decrement values for the Balmer lines. The corrections should find future use with integrated-light models to better predict stellar populations parameters, especially the mean metallicities and ages of galaxies.

Our analysis methods and results are set out in the following section with a discussion and summary in $\S$ 3.

\section{Analysis and Results}

The first panel of Figure 1 shows that there is a discrepancy in the H$\alpha$ measurements between the SDSS and Virgo spectra for this work use measurements of the H$\alpha$ index as defined by \cite{Cohen}. Plausibly, this is either due to hydrogen emission in the SDSS spectra or to nonspectrophotometric wavelength-dependent fluxing errors that would propagate into Lick index measurements depending on how the various passbands lay across the various spectral ``lumps.'' We infer that the bulk of the discrepancy is most likely due to hydrogen emission in the SDSS spectra because the continuum-shape differences between the data sets is too small to account for much of the discrepancy, as we now show.

To determine if differences in the continuum of the two data sets could explain this discrepancy, the ratio of similar spectra (similar as regards velocity dispersion) from the two data sets was computed and a continuum fit was found for this ratio using the ``continuum'' routine in IRAF \footnote{IRAF is distributed by the National Optical Astronomy Observatories, which are operated by the Association of Universities for Research in Astronomy, Inc., under Cooperative agreement with the National Science Foundation.}. Lick style indices were measured from the  {\em fit} as one would measure spectra. The delta-index thus discovered was found to be small: about 0.06\AA\ of the measured values for H$\alpha$, insufficient to explain the SDSS-Virgo discrepancy. It should be noted that the effects of the continuum differences {\em could} be more severe if differences in the continuum shapes exist that are of the same wavelength span as the features of interest. Most would likely lie with the SDSS spectra, since our fluxing of the Virgo data was careful, but, unfortunately we don't have the data to check the significance of these differences.

We characterized the SDSS-Virgo H$\alpha$ - Mg $b$ trend by best fit lines calculated using fitexy.f \citep{numrec}. This is a fortran program for finding the best fit line for data with errors in both the x and y coordinate. This minimizes the distance of each point form the line while taking into account weighting by the precision of the individual measurements in both the x and y coordinate. The choice to characterize the trends in H$\alpha$ as a function of Mg $b$ was for two reasons. One was the tight correlation between Mg $b$ and $\sigma$ and the other was for when using models as described later Mg $b$ was just practical. Those line fits and the root mean squares (RMS) of the distances of the points from their fit lines. Also shown in Table 1 is the form of the correction term (${{ j_\alpha }\over{F_{c,\alpha}}}$) determined from the following derivation.

\begin{table}[H]
\begin{center}
\begin{tabular}{ l c c }
\multicolumn{3}{c}{Table 1} \\
\hline
\hline
Data set & Line Fit in \AA\ & RMS of Fit in \AA\ \\
\hline
Virgo      & $3.0132-0.3768 *$Mg $b$ & $0.066$ \\
SDSS       & $1.3476-0.0532 *$Mg $b$ & $0.020$ \\
Correction Term  &$1.6656-0.3236 *$Mg $b$ $= {{ j_\alpha }\over{F_{c,\alpha}}}$ & $0.069$ \\
\hline
\end{tabular}
\caption{Shown in this table are the best fit H$\alpha$ vs. Mg $b$ lines for the Virgo and Sloan data sets and their associated RMS values (lines 1 and 2). Line 3 is the difference between these two best fit lines, which represents the H$\alpha$ correction term along with it's RMS value.}
\end{center}
\end{table}

Using the linear correction term for the hydrogen emission in H$\alpha$ a correction for the subsequent Balmer lines was constructed under the assumption that the entire shift is due to Balmer emission fill-in. To determine the form of the correction term we start with the definition of equivalent width (EW; see Eq. 1). In Eq. 1, $\mathbf{\lambda_{1}} $ and $\mathbf{\lambda_{2}} $ are defined as the blue and red wavelength bounds of the index passband and F$_{i}$ and F$_{c}$ are the average index flux and the pseudocontinuum flux respectively see \cite{worthey94}.

\begin{equation}
  EW=  \int_{\lambda 1}^{\lambda 2}   ( 1 - {{F_\lambda }\over{F_c}} ) d\lambda = \Delta \lambda ( 1 - {\overline{F_\lambda}\over{F_c}} )   
  \end{equation}

With more generality, and including a term for the flux due to an emission feature,

\begin{equation}
  EW=  \int_{\lambda 1}^{\lambda 2}   ( 1 - {{F_\lambda + F_j}\over{F_c}}   ) d\lambda = \Delta \lambda ( 1 - {\overline{F_\lambda}\over{F_c}} - {\overline{F_j}\over{F_c}} )   
  \end{equation}

where $F_j$ is the flux of the emission feature and $F_\lambda$ is the flux of the stellar light and $\Delta \lambda = \lambda_2 - \lambda_1$. One should note, however, that it is the emission line's power that is constant. If we call the line's power $j$ then the average emission line flux is defined by

\begin{equation}
 j = \int_{\lambda 1}^{\lambda 2} F_j d\lambda = \Delta \lambda  \overline{F_j}
\end{equation}

Thus, the average stellar flux inside the continuum band is $F_i = \overline{F_\lambda}$ and the correction term for the equivalent width is $\Delta \lambda \overline{F_j}/F_c =j / F_c$.  To extend to Balmer lines other than H$\alpha$, we exploit the fact that the decrements $j_\beta/j_\alpha$, $j_\gamma/j_\alpha$, and $j_\delta/j_\alpha$ are known, and relatively constant.

For example, if $j_\alpha$ is known, then extending from an H$\alpha$ index to an H$\beta$ index is accomplished by adding the correction term 

\begin{equation}
{{ j_\beta }\over{F_{c,\beta}}} = {{ j_\alpha }\over{F_{c,\alpha}}}  \times {{F_{c,\alpha}}\over{F_{c,\beta}}} {{j_\beta}\over{j_\alpha}}
\end{equation}

Having determined the H$\alpha$ correction from best line fits (see Table 1) all that is left to do is determine the conversion factors  ${{F_{c,\alpha}}\over{F_{c,\beta}}}$ and ${{j_\beta}\over{j_\alpha}}$.

For this paper a version of the \cite{wor94} and  \cite{trag}) models were used that use a grid of synthetic spectra in the optical \cite{lee} in order to investigate the effects of changing the detailed elemental composition on an integrated spectrum was used to create synthetic spectra at a variety of ages and metallicities for single-burst stellar populations. The underlying isochrones for this paper were the \cite{wor94} isochrones, because they allow us "manual" HB morphology control. However, there are certain caveats to using these isochrones. specifically the models are a bit crude by today's standards and the ages are about 2 Gyr too old, so that 17 Gyr should really be interpreted as 15 Gyr. Other isochrone sets were used to check the results.

For this paper new index fitting functions were generated. The data sources include a variant of the original lick collection of stellar spectra \citep{worthey94} in which the wavelength scale of each observation has been refined via cross-correlation, as well as the MILES spectral library \citep{san} with some zero point corrections, and the Coude Feed Library (CFL) of \cite{valdes}. The CFL was used as the fiducial set, in the sense that any zero point shifts between libraries were corrected to agree with the CFL case. The MILES and CFL spectra were smoothed to a common Gaussian smoothing corresponding to 200 km $s^{-1}$. The rectified-Lick spectra were measured and then a linear transformation was applied to put it on the fiducial system.

Multivariate polynomial fitting was done in five overlapping temperature swaths as a function of $\Theta$$_{eff}$ = 5040/T$_{eff}$, log g, and [Fe/H]. The fits were combined into a lookup table for final use. As in \cite{wor94}, an index was looked up for each "star" in the isochrone and decomposed into "index" and " continuum" fluxes, which added, then re-formed into an index representing the final, integrated value after the summation. This gives us empirical synthetic spectra when variations in chemical composition are needed. The grid of synthetic spectra is complete enough to predict nearly arbitrary composition.

These models were then used in order to determine the conversion factors for the continuum levels. By first taking the ratio of the continuum levels near H$\alpha$ and the rest of the Balmer lines as measured from these models and plotting them against Mg $b$. Then, fitting a least squares line to the data. This line then gives the change of the relative continuum levels as a function of Mg $b$. 

The last piece of the puzzle is that the native relative intensities of the Balmer lines need to be accounted for. This is taken into account by the relative Balmer line intensities as calculated by \cite{oster}. We adopt case B, 10000 K conditions, because they are near the middle of the range for star formation regions ($j_\alpha/j_\beta = 2.85$), and they are not too drastically different than LINER type spectra j$_\alpha/j_\beta = 3.27$ from \cite{graves07}. The final correction formulae were applied to H$\beta$ defined in \cite{worthey94}, H$\gamma_A$, H$\gamma_F$, H$\delta_A$, and H$\delta_F$ as defined in \citep{worthey97} and are found in Table 2.

\begin{table}[H]
\begin{center}
\begin{tabular}{l c c c c c}
\multicolumn{6}{c}{Table 2} \\
\hline
\hline
Balmer&Correction&H$\alpha$ Correction&Continuum Correction&Decrement&Correction term \\
Index& Term in \AA\ & in \AA\ (${{ j_\alpha }\over{F_{c,\alpha}}}$)& in \AA\ (${{F_{c,\alpha}}\over{F_{c,i}}}$)&(${{j_i}\over{j_\alpha}}$)& RMS of Fit in \AA\ \\
\hline
H$\beta$&${{ j_\beta }\over{F_{c,\beta}}}=$   &$(1.666-0.324$(Mg $b$)) $\times$& $(0.838+0.076$(Mg $b$)) $\times$& $0.351$ & $0.074$ \\
H$\gamma_A$&${{ j_\gamma }\over{F_{c,\gamma}}}=$  &$(1.666-0.324$(Mg $b$)) $\times$& $(0.847+0.149$(Mg $b$)) $\times$& $0.165$ & $0.091$ \\
H$\gamma_F$&${{ j_\gamma }\over{F_{c,\gamma}}}=$ &$(1.666-0.324$(Mg $b$)) $\times$& $(0.847+0.149$(Mg $b$)) $\times$& $0.165$&$0.091$ \\
H$\delta_A$&${{ j_\delta }\over{F_{c,\delta}}}=$  &$(1.666-0.324$(Mg $b$)) $\times$& $(0.649+0.293$(Mg $b$)) $\times$& $0.091$ & $0.139$ \\
H$\delta_F$&${{ j_\delta }\over{F_{c,\delta}}}=$   &$(1.666-0.324$(Mg $b$)) $\times$& $(0.649+0.293$(Mg $b$)) $\times$& $0.091$ & $0.139$ \\
\hline
\end{tabular}
\caption{The range over which Mg $b$ yields the most reliable results is 3.0 \AA\ $<$ Mg $b$ $<$ 4.3 \AA\ . This Mg $b$ range covers most elliptical galaxies.}
\end{center}
\end{table}

\begin{figure}[H]
\includegraphics[width=3in]{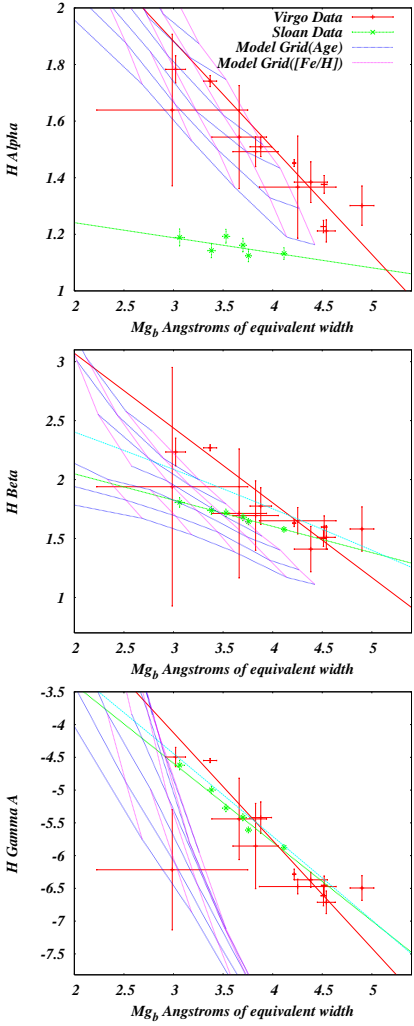}
\caption{H$\alpha$, H$\beta$,and H$\gamma_A$ against Mg $b$ indices for two data sets and models. The Virgo spectra (red symbols with error bars), the SDSS spectra (green symbols with error bars), single stellar population models (blue grid lines) from bottom to top of ages 17, 12, 8, 5, 3, 2, and 1.5 Gyrs and models from right to left of metallicites $0.5, 0.25, 0, -0.25, -0.5, -1, -1.5,$ and $-2$ (pink grid lines) are plotted. The fits to the index values  for the SDSS galaxies after the correction for hydrogen emission are shown as a light blue lines in each panel.
\label{fig1} }
\end{figure}

Figures 1 and 2  show the measurements of Virgo and SDSS galaxy averages along with the corrected SDSS data (light blue line). In all the graphs the Virgo data is in red while the SDSS data is in green. Also in Figures 1 and 2 model grids are plotted in blue and pink. The blue corresponds to models of various ages from 1.5 to 17 Gyrs and the pink corresponds to models of various metallicities from -2.00 to 0.50. For the H$\beta$ correction the line fit is a little low for smaller galaxies with weaker Mg $b$, but not bad for the larger galaxies. See Figure \ref{fig1}.

% Fig 1 here

The fits for H$\gamma$ and H$\delta$ do not look as good when compared to the Virgo measurements (c.f. Figure 2). The reason is most likely the sensitivity of the $H\gamma$ and $H\delta$ indices to changes in individual elemental abundances, with perhaps a non-negligible contribution from systematic errors between the two data sets. Elements N, C, and O have a profound and interdependent effect on this region of spectrum, and suggest themselves as candidates for further investigation.

%Fig 2 here
\begin{figure}[H]
\includegraphics[width=3in]{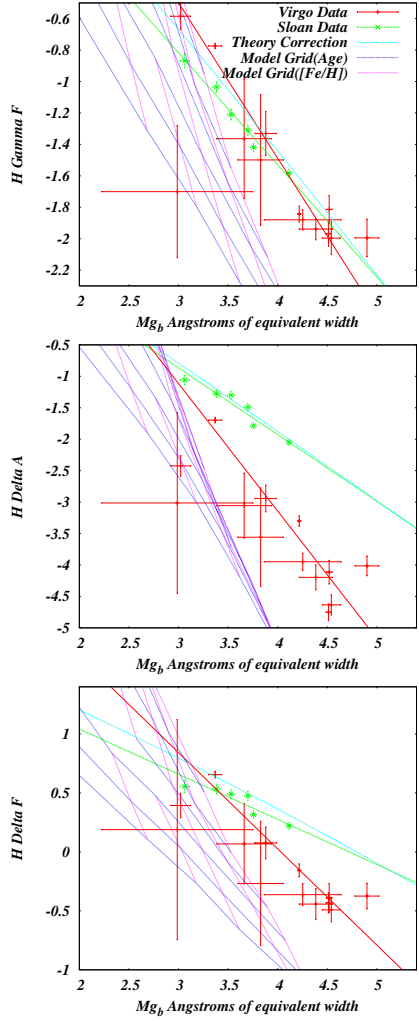}
\caption{H$\gamma_F$, H$\delta_A$,and H$\delta_F$ against Mg $b$ indices for two data sets and models. The Virgo spectra (red symbols with error bars), the SDSS spectra (green symbols with error bars), single stellar population models (blue grid lines) from bottom to top of ages 17, 12, 8, 5, 3, 2, and 1.5 Gyrs and models from right to left of metallicites $0.5, 0.25, 0, -0.25, -0.5, -1, -1.5,$ and $-2$ (pink grid lines) are plotted. The fits to the index values  for the SDSS galaxies after the correction for hydrogen emission are shown as a light blue lines in each panel.
\label{fig2}}
\end{figure}

\section{Discussion, Summary, and Conclusion}

Using the correction factor (${{ j_\beta }\over{F_{c,\beta}}}$) as derived above and associating it with the linear offset in H$\alpha$ this work has shown that for SDSS spectra cleaned like those in \cite{graves07}, a emission correction factor on the order of  0.5 \AA\ in H$\alpha$ and 0.2 \AA\ in H$\beta$ needs to be applied to the quiescent galaxies in order to better determine the mean age and metallicity of these galactic averages. These correction factors are much larger than those estimated in \cite{graves07}, where the estimated correction factors from measurements of O{\sc II} give H$\alpha$ = 0.082 \AA\ and H$\beta$ = 0.027 \AA\/. This work also shows  that the higher-order Balmer lines H$\gamma$ and H$\delta$ have such modest emission corrections that  other uncertainties dominate the error budget.  At least for our comparison sample of Virgo cluster elliptical galaxies, H$\gamma$ and H$\delta$ suffer from contamination from the varying of other elemental abundances, making age determination from these indices more complicated. This makes the need for correct H$\alpha$ and H$\beta$ measurements all the more important, since they are relatively insensitive to changes in metal abundances.

These results are in qualitative agreement with the results from \cite{eisen} who showed that in their SDSS spectra H$\beta$ suffers from interstellar emission lines and speculated that non-solar abundance ratios were to blame for the differences in age determinations made from H$\beta$, H$\gamma$ and H$\delta$. This is in agreement with these findings, where not only is there obvious hydrogen emission, but also it is clear that relative abundance ratios would certainly have to be taken into account in order to determine galactic ages from H$\gamma$ and H$\delta$.

For this work the H$\alpha$ correction was chosen as the basis for the corrections due to the facts that H$\alpha$ is 3 times more sensitive to hydrogen emission than H$\beta$ and the line fit for H$\alpha$ as measured in the virgo galaxies had a much tighter fit RMS = 0.066 \AA\ compared with H$\beta$ RMS = 0.15 \AA\ leading to a corrected SDSS line fit with much less uncertainty. Especially  for galaxies inside the range 3.0 $<$ Mg $b$ $<$ 4.3 \AA\ where the corrections for H$\beta$ yield the most reliable results. Another reason for choosing H$\alpha$ instead of directly using the correction that one could get from the H$\beta$ plot is that we wanted to preserve H$\beta$ for age determination.

The plot for H$\beta$ in figure 1 panel 2 still shows a discrepancy between the corrected SDSS data (light blue line) and the Virgo data (red line). This residual discrepancy could be due to a few variables such as varying abundance ratios within the virgo galaxies, slightly different decrements values than the ones used here, or there is the possibility that the H$\alpha$ emission correction may be in part due to an difference in the mean ages of the two samples. \cite{thomas05} showed that cluster galaxies tend to be around 2 Gyrs older than field galaxies.

Another possibility is that H$\alpha$ and H$\beta$ have different age sensitivities. In order to investigate the age sensitivities using the aforementioned models the Z sensitivity parameter was calculated for both H$\alpha$ and H$\beta$ as it was in \cite{worthey94}. The age sensitivity parameter is the ratio of the percentage change in age to the percentage change in Z of the index measured as shown below. 

\begin{equation}
 Z = {[\Delta I/(\Delta Z/Z)]\over[\Delta I / (\Delta age/age)]}
\end{equation}
 
Here $\Delta$I /($\Delta$Z/Z) is the average of the changes in the measured index I as measured from the models where the age is 12 Gyrs and [Fe/H] changes form [Fe/H] = 0.25 to [Fe/H] = 0.0 and [Fe/H] = 0.0 to [Fe/H] = -0.25 and Z= 0.0169 $\times$ $10^{[Fe/H]}$. Similarly, $\Delta$I /($\Delta$age/age) is the average of the changes in the measured index I as measured from the models where the metallicity is solar and the age changes from 17 Gyrs to 12 Gyrs and 12 Gyrs to 8 Gyrs and age = 12 Gyrs. These sensitivities are shown in Table 3 along with the original \cite{worthey94} H$\beta$ sensitivity. Note that the models indicate that both H$\alpha$ and H$\beta$ are age indicators of the same sensitivity. This means that the discrepancy between H$\alpha$ and H$\beta$ is more likely to be due to abundance ratios or the decrement values used than it is the age sensitivities of H$\alpha$ and H$\beta$.  

\begin{table}[H]
\begin{center}
\begin{tabular}{ l c }
\multicolumn{2}{c}{Table 3} \\
\hline
\hline
Index & Z (sensitivity parameter) \\
\hline
H$\alpha$      &  0.8 \\
H$\beta$      &  0.8 \\
H$\beta$ (Worthey et. al 94)& 0.6\\
\hline
\end{tabular}
\end{center}
\end{table}

The applicability of this work for other grand SDSS averages may be limited due to the details of the sample selection. It is also unlikely to be applicable to individual red sequence galaxies due to the wide dispersion in index values and possible age effects. One possible application could be to scale those averages to match those of \cite{graves07}, but it would be a better idea to echo the work shown here by comparing with the minimal-emission Virgo data set those averages and determining a new correction.

It is conceptually possible to solve for both age and emission correction by considering both H$\beta$ and H$\alpha$ simultaneously, and increasing the emission correction until both indices give similar ages against a model grid. The obvious trouble with that is that the solution then becomes model dependent. There is also strong anticorrelation between derived age and emission correction. Finally, there is the contribution of N emission near H$\alpha$, whose presence may cause spuriously large H$\alpha$ emission measurements in individual galaxies.

Speculation aside, it is safe to say that there is emission contamination in the SDSS spectra and that it is reasonably well accounted for by the linear fits presented in this paper. In data sets  that include H$\alpha$, observed OII and OIII do not need to be used as a proxy for Balmer emission.  Emission contamination is much less of a problem for H$\gamma$ and H$\delta$, but interpretation of these indices is complicated by the probable effects of individual elemental abundances.

\acknowledgements
We would like to thank Genevieve J. Graves for providing the Sloan Digital Sky Survey spectra, as well as her advice and input for this paper. Major funding for this work was provided by National Science Foundation grants 0307487 and 0346347.

\end{document}